\documentclass[journal]{IEEEtran}

\usepackage{amsmath}
\usepackage{amssymb}
\usepackage{graphicx}
\usepackage{booktabs}
\usepackage{multirow}
\usepackage{cite}
\usepackage{url}
\usepackage{hyperref}
\usepackage{array}
\usepackage{textcomp}
\usepackage{float}

\hypersetup{
    colorlinks=true,
    linkcolor=black,
    citecolor=black,
    urlcolor=blue,
    pdfauthor={Andrii Vakhnovskyi},
    pdftitle={IOGRUCloud: A Scalable AI-Driven IoT Platform for Climate Control in Controlled Environment Agriculture}
}

\begin{document}

\title{IOGRUCloud: A Scalable AI-Driven IoT Platform for Climate Control in Controlled Environment Agriculture}

\author{%
\IEEEauthorblockN{ANDRII VAKHNOVSKYI}\\%
\IEEEauthorblockA{IOGRU LLC, New York, NY 10022, USA}%
\thanks{Corresponding author: Andrii Vakhnovskyi (\mbox{e-mail:} \mbox{andrii.vakhnovskyi@gmail.com}). ORCID: 0009-0007-8306-5932.}%
}

\maketitle

\begin{abstract}
Climate control in Controlled Environment Agriculture (CEA) remains dominated by static-setpoint controllers with isolated PID loops, leading to cross-coupling conflicts, suboptimal energy consumption, and limited scalability across heterogeneous facilities. This paper presents IOGRUCloud, a three-tier IoT platform for AI-driven climate control deployed across 30+ commercial CEA facilities in 8 U.S. climate zones over 7+ years of continuous operation (2017--2024) --- the largest documented real-world deployment of adaptive climate control in this domain, exceeding the combined duration of all published field experiments in AI-based HVAC control by approximately 60 times. The platform introduces a cascading control architecture where Vapor Pressure Deficit serves as the primary setpoint, with a neural network optimizer selecting energy-minimal temperature--humidity combinations on the VPD constraint surface. Inner PID loops with neural network self-tuning track the resulting setpoints at higher bandwidth. A four-level progressive autonomy model (L1--L4) enables graduated transition from anomaly detection to autonomous optimization with operator guardrails. The edge-first architecture ensures all control logic executes locally without cloud dependency, while the cloud tier enables cross-facility transfer learning across the fleet. Multi-facility aggregate results demonstrate 30--38\% HVAC energy reduction, 68--73\% improvement in VPD stability, and 60--67\% faster disturbance recovery compared to conventional controllers. Two detailed case studies --- a 40,000~sq~ft desert-climate facility and a 120,000~sq~ft continental-climate facility --- are presented with quantitative metrics. Practical deployment lessons from integrating 50+ equipment manufacturers via 8 industrial protocols and achieving 1--5 day commissioning timelines are discussed.
\end{abstract}


\section{Introduction}

\IEEEPARstart{T}{HE} global Controlled Environment Agriculture (CEA) industry is undergoing rapid expansion as climate change, urbanization, and food security concerns drive demand for year-round, location-independent crop production~\cite{ref1, ref2}. CEA facilities --- including greenhouses, vertical farms, and indoor cultivation operations --- require precise simultaneous management of dozens of interdependent environmental parameters: air temperature, relative humidity, CO\textsubscript{2} concentration, photosynthetically active radiation (PAR), substrate moisture, and nutrient solution chemistry. Energy expenditure for climate regulation accounts for 20--50\% of CEA operating budgets, with HVAC systems alone consuming 30--80\% of total facility energy depending on geographic location and outdoor climate conditions~\cite{ref3, ref4}.

Conventional CEA climate control relies on static-setpoint controllers with isolated PID loops for individual parameters~\cite{ref5}. This approach suffers from fundamental limitations: (1)~static setpoints do not adapt to changing outdoor conditions, leading to suboptimal energy consumption; (2)~independent PID loops create cross-coupling conflicts --- for example, simultaneous heating and dehumidification operations that work against each other; (3)~reactive control responds to deviations after they occur rather than anticipating them; and (4)~the lack of cross-subsystem coordination results in energy waste from contradictory actuator commands~\cite{ref6}.

Recent advances in artificial intelligence have shown promise for CEA climate control. Chen~\textit{et~al.}~\cite{ref7} review AI applications in CEA, identifying deep reinforcement learning and model predictive control as leading approaches. Ajagekar~\textit{et~al.}~\cite{ref8} demonstrated up to 57\% energy reduction using deep reinforcement learning with robust optimization --- however, this result was obtained in simulation on a single greenhouse model. Adesanya~\textit{et~al.}~\cite{ref9} applied deep reinforcement learning for PID parameter tuning in greenhouse HVAC, while Panagopoulos~\textit{et~al.}~\cite{ref10} proposed a cascaded economic model predictive control approach for greenhouse climate --- both validated only in simulation. The iGrow system~\cite{ref11} demonstrated reinforcement learning for autonomous greenhouse control with measurable yield improvements, but at limited scale.

The hybrid approach combining classical PID control with neural network adaptation has been explored since the work of Zeng~\textit{et~al.}~\cite{ref12}, who proposed RBF neural network-augmented PID for greenhouse temperature regulation. Salehi~\textit{et~al.}~\cite{ref13} validated neural network-based PID auto-tuning in an industrial setting, demonstrating production-scale reliability. The concept of progressive industrial autonomy has been formalized by Gamer~\textit{et~al.}~\cite{ref14}, who proposed a six-level taxonomy for autonomous industrial plants. The IoT and edge-cloud computing paradigm for agricultural applications has been established through reference architectures by Alreshidi~\cite{ref15} and Sami and Ibraheem~\cite{ref16}. The physiological basis for using Vapor Pressure Deficit (VPD) as a primary control variable is well-established: Grossiord~\textit{et~al.}~\cite{ref17} demonstrated that VPD is a critical driver of plant stress responses, while Inoue~\textit{et~al.}~\cite{ref18} showed that minimizing VPD fluctuations directly improves plant growth.

Despite these advances, a critical gap persists between simulation-validated research and production-scale deployment. Mulayim~\textit{et~al.}~\cite{ref19} found that the combined duration of all peer-reviewed real-world field experiments in AI-based HVAC control totals approximately 43~days globally. Al~Sayed~\textit{et~al.}~\cite{ref20} reported that only 23\% of reinforcement learning studies for HVAC involved real buildings. The largest published real-world deployment --- Moshari~\textit{et~al.}~\cite{ref21} with 13~buildings over one heating season --- remains limited in both facility count and temporal scope.

Furthermore, no commercial CEA control platform --- including Priva, Argus Controls, TrolMaster, or Wadsworth --- has published its system architecture, control algorithms, or deployment-scale performance data in peer-reviewed literature~\cite{ref5}. The only comparable end-to-end agricultural IoT platform paper is FarmBeats~\cite{ref22}, which describes a monitoring and analytics platform (not a closed-loop control system) deployed at 2~farms.

This paper addresses these gaps. The main contributions are as follows:

\begin{enumerate}
\item We present the architecture of IOGRUCloud, a three-tier IoT platform for CEA climate control deployed across 30+ commercial facilities in 8 U.S. climate zones --- the largest peer-reviewed deployment of its kind, exceeding all published agricultural IoT field experiments by an order of magnitude.

\item We propose a cascading VPD control architecture where Vapor Pressure Deficit serves as the primary setpoint, and a neural network optimizer determines the energy-minimal temperature and humidity targets on the VPD constraint surface.

\item We describe a neural network-based PID self-tuning mechanism with Lyapunov stability guarantees, deployed in production HVAC systems --- the first reported real-world deployment of neural network-tuned PID controllers in CEA.

\item We introduce a four-level progressive autonomy model (L1--L4) for CEA automation, with integrated confidence scoring and operator guardrails.

\item We report production deployment results including 30--38\% HVAC energy reduction and practical lessons from integrating 50+ equipment manufacturers via 8 industrial protocols with 1--5 day commissioning timelines.
\end{enumerate}

The remainder of this paper is organized as follows. Section~II describes the three-tier system architecture. Section~III presents the cascading VPD control approach and neural network PID self-tuning with stability analysis. Section~IV introduces the progressive autonomy model. Section~V describes the multi-facility deployment methodology. Section~VI presents experimental results from 30+ facilities with two detailed case studies. Section~VII discusses practical deployment lessons. Section~VIII concludes the paper.

\section{System Architecture}

The IOGRUCloud platform implements a three-tier architecture with clearly separated responsibilities, illustrated in Fig.~1.

\begin{figure}[!t]
\centering
\includegraphics[width=\columnwidth]{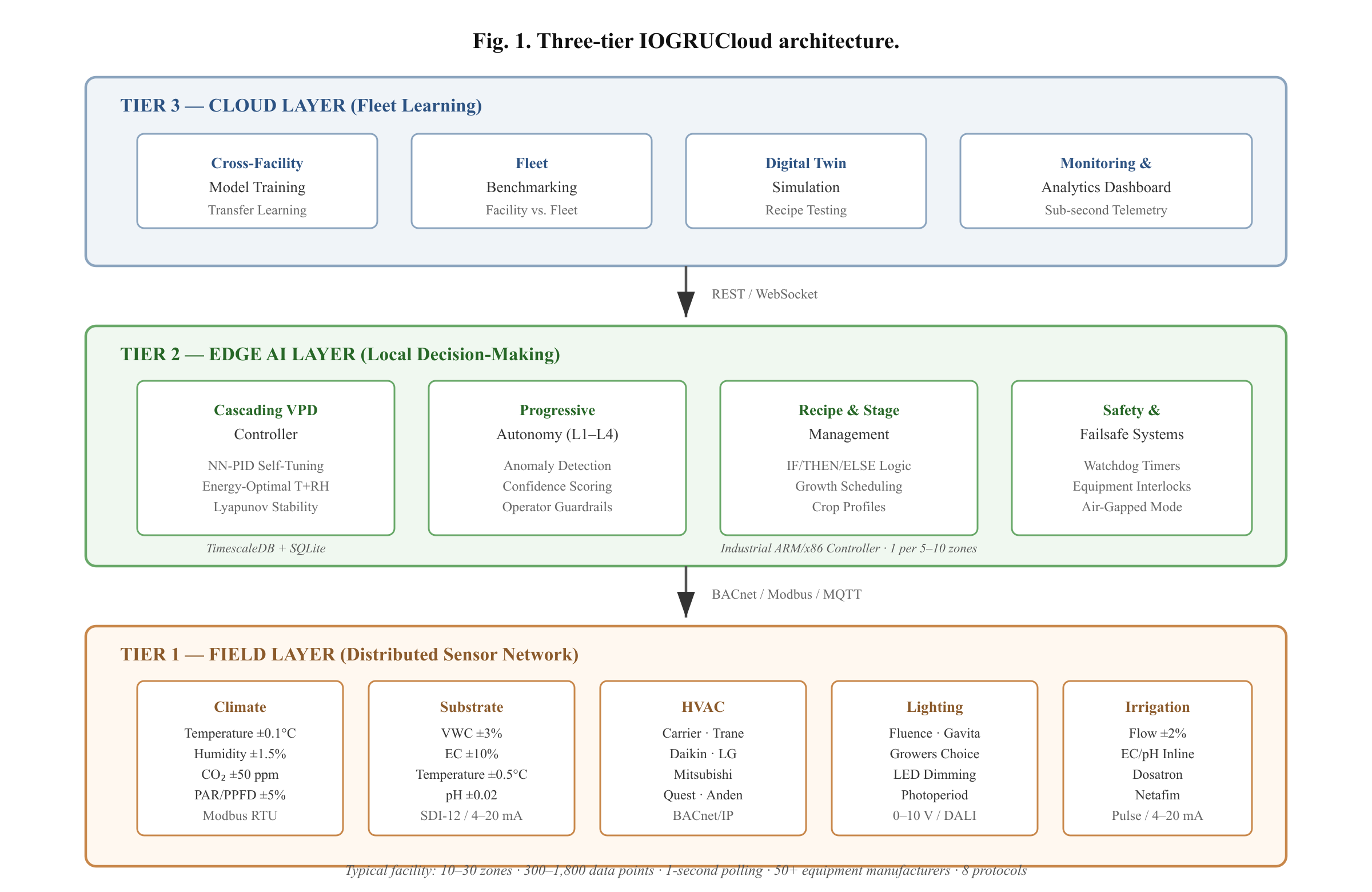}
\caption{Three-tier architecture diagram: Field Layer $\rightarrow$ Edge AI Layer $\rightarrow$ Cloud Layer.}
\label{fig:architecture}
\end{figure}

\subsection{Field Layer (Distributed Sensor Network)}

The field layer comprises a distributed network of industrial sensors and actuators communicating via standard protocols. Table~\ref{tab:sensors} lists the sensor types and specifications deployed in a typical CEA facility zone.

\begin{table*}[!t]
\centering
\caption{Sensor Specifications Per Control Zone}
\label{tab:sensors}
\begin{tabular}{@{}lllll@{}}
\toprule
\textbf{Parameter} & \textbf{Sensor type} & \textbf{Accuracy} & \textbf{Interface} & \textbf{Qty/zone} \\
\midrule
Air temperature & Aspirated climate station & $\pm 0.1\,^{\circ}$C & Modbus RTU & 2--4 \\
Relative humidity & Aspirated climate station & $\pm 1.5\%$ RH & Modbus RTU & 2--4 \\
CO\textsubscript{2} concentration & NDIR sensor & $\pm 50$~ppm & Modbus / 4--20~mA & 1--2 \\
PAR/PPFD & Quantum sensor (400--700~nm) & $\pm 5\%$ & SDI-12 / 4--20~mA & 1--2 \\
Substrate VWC & Capacitance probe & $\pm 3\%$ & SDI-12 & 3--8 \\
Substrate EC & Capacitance probe & $\pm 10\%$ & SDI-12 & 3--8 \\
Substrate temperature & Thermistor & $\pm 0.5\,^{\circ}$C & SDI-12 & 3--8 \\
Solution EC (in/out) & Inline transmitter & $\pm 2\%$ & 4--20~mA / Modbus & 2 \\
Solution pH (in/out) & ISFET / glass electrode & $\pm 0.02$ & 4--20~mA / Modbus & 2 \\
Water flow & Pulse flow meter & $\pm 2\%$ & Pulse & 1--4 \\
Differential pressure & Transmitter & $\pm 1$~Pa & 4--20~mA & 1 \\
Power consumption & CT clamp per circuit & $\pm 1\%$ & 4--20~mA & 4--16 \\
Leaf temperature & IR thermometer & $\pm 0.5\,^{\circ}$C & Modbus & 1--2 \\
\bottomrule
\end{tabular}
\end{table*}

A typical multi-zone facility (10--30 zones) generates 300--1,800 data points at 1-second polling frequency, with storage at 10-second resolution. Large multi-state operator facilities generate up to 3,000+ data points. The platform integrates equipment from 50+ manufacturers --- including Carrier, Trane, Daikin, LG, and Mitsubishi for HVAC; Quest, Anden, and Desert Aire for dehumidification; and Fluence, Gavita, and Growers Choice for lighting --- through a vendor-agnostic integration layer spanning 8 industrial protocols: BACnet/IP, Modbus RTU/TCP, MQTT, OPC~UA, 0--10~V analog, 4--20~mA analog, SDI-12, and REST/WebSocket APIs.

Multi-sensor redundancy within each zone employs median filtering across co-located sensors of the same type to detect and reject sensor drift without corrupting the aggregate measurement~\cite{ref23}. When the z-score of any individual sensor reading exceeds 2.5 relative to its co-located peers, the reading is flagged for maintenance review and excluded from the control signal.

\subsection{Edge AI Layer (Local Decision-Making)}

The edge controller is the central architectural element. All real-time control logic executes locally on industrial ARM/x86 hardware without cloud dependency --- a critical requirement for commercial agricultural facilities where network outages must not disrupt climate control.

The edge layer hosts:

\begin{enumerate}
\item The cascading VPD controller with PID auto-tuning (Section~III).
\item The neural network AI module implementing progressive autonomy levels L1--L4 (Section~IV).
\item Recipe management and stage-based automation with configurable IF/THEN/ELSE logic.
\item Safety systems including hardware watchdog timers, failsafe modes, and equipment interlock logic.
\end{enumerate}

Local data storage uses TimescaleDB for time-series data at 10-second resolution (supporting 2--8 years of retention) and SQLite for configuration. The air-gapped failsafe mode ensures continued operation during network outages: the edge controller maintains full autonomous control capability using locally cached recipes, setpoints, and learned parameters. Typical edge configuration: 1~controller per 5--10 zones.

\subsection{Cloud Layer (Fleet Learning)}

The cloud tier aggregates anonymized operational data from the facility network and enables capabilities that require cross-facility visibility:

\begin{enumerate}
\item \textit{Cross-facility model training} --- learned operational patterns from established facilities are used to accelerate optimization at new deployments. Optimal VPD trajectories for specific crop stages, equipment-specific PID tuning parameters, and seasonal control strategy templates are transferred through the fleet~\cite{ref24}.
\item \textit{Multi-facility benchmarking} --- facilities operating similar crops under comparable recipes are compared to identify optimization opportunities.
\item \textit{Digital twin simulation} --- proposed recipe modifications are simulated against historical data before application to live facilities~\cite{ref25}.
\item \textit{Real-time monitoring and alarming} --- sub-second telemetry across all deployed sites with centralized operational analytics.
\end{enumerate}

The cloud layer supports L4-level autonomous optimization (Section~IV), including predictive yield modeling and energy cost optimization via load shifting to off-peak tariff hours.

\section{AI-Driven Climate Control}

\subsection{Cascading VPD Control Architecture}

The central innovation in control strategy is the elevation of VPD from a monitored metric to the primary cascading control variable. Rather than independently regulating temperature and humidity --- which often produces contradictory actuator commands --- the system targets VPD directly and uses a neural network optimizer to decompose VPD targets into energy-optimal temperature--humidity combinations.

The cascading architecture operates as follows (Fig.~2):

\begin{figure}[!t]
\centering
\includegraphics[width=\columnwidth]{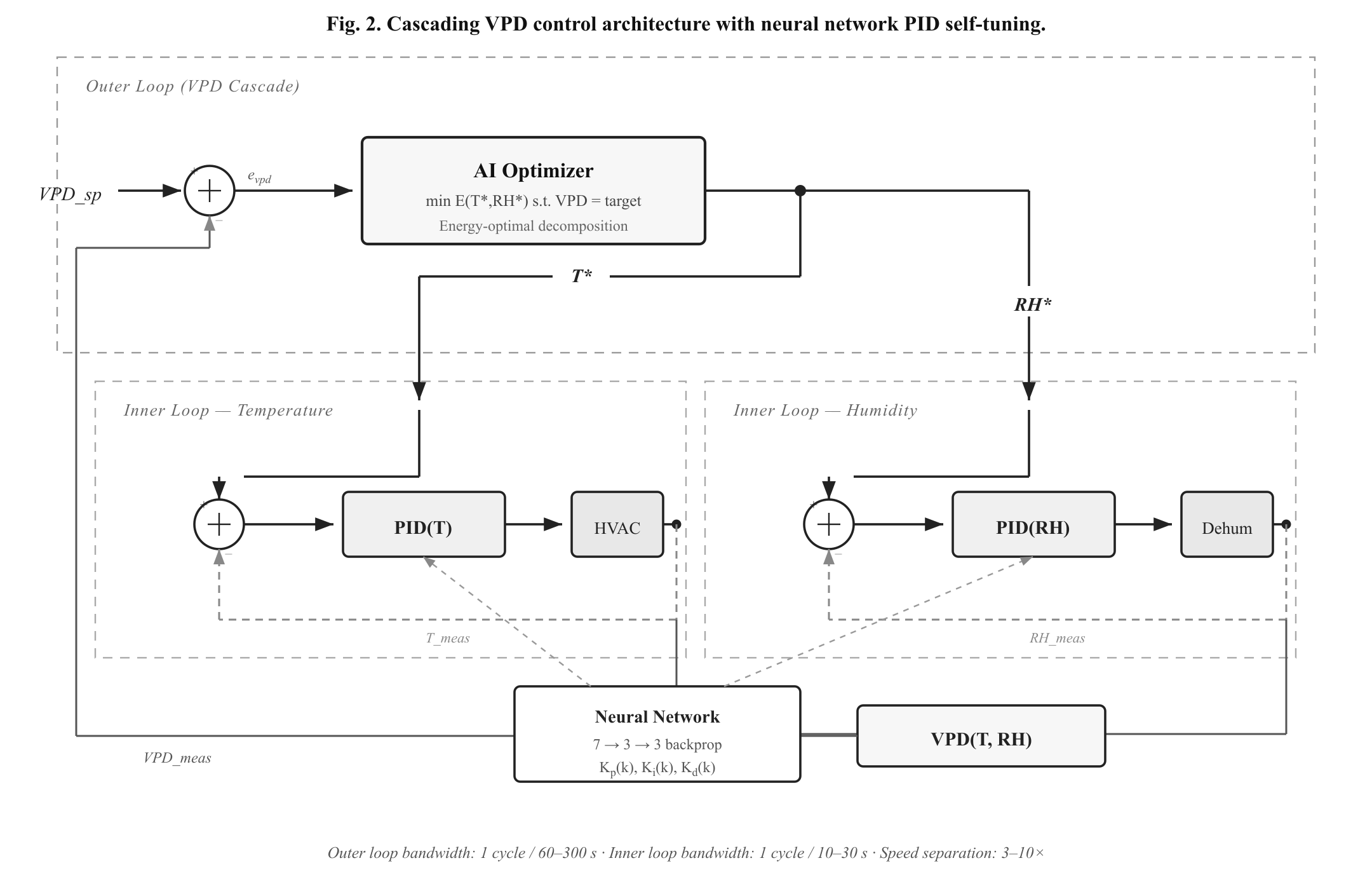}
\caption{Cascading VPD control block diagram.}
\label{fig:vpd_cascade}
\end{figure}

The outer loop (AI optimizer) operates on VPD error and selects the energy-minimal point on the VPD constraint surface in $T$--$RH$ space. The inner loops (PID controllers) track the resulting temperature and humidity setpoints. The inner loops operate at 3--10$\times$ higher bandwidth than the outer loop, satisfying the cascade speed separation requirement~\cite{ref10}.

The saturation vapor pressure is computed using the Tetens equation with improved coefficients from Alduchov and Eskridge~\cite{ref26}:

\begin{equation}
e_s(T) = 0.61094 \times \exp\!\left(\frac{17.625 \times T}{T + 243.04}\right)
\label{eq:tetens}
\end{equation}

where $e_s$ is in kPa and $T$ in $^{\circ}$C. This improved Magnus form yields relative errors below 0.01\% in the 0--50$\,^{\circ}$C range.

The Vapor Pressure Deficit is:

\begin{equation}
\text{VPD} = e_s(T_{\text{air}}) \times \left(1 - \frac{\text{RH}}{100}\right)
\label{eq:vpd}
\end{equation}

For facilities equipped with infrared canopy temperature sensors:

\begin{equation}
\text{VPD}_{\text{leaf}} = e_s(T_{\text{leaf}}) - e_s(T_{\text{air}}) \times \frac{\text{RH}}{100}
\label{eq:vpd_leaf}
\end{equation}

The system supports three VPD computation modes: air VPD (standard), leaf VPD (with canopy sensors), and canopy VPD (multi-point spatial average)~\cite{ref17, ref27}.

\subsection{Energy-Optimal VPD Decomposition}

The VPD constraint $e_s(T^*) \times (1 - \text{RH}^*/100) = \text{VPD}_{\text{target}}$ defines a curve in $T$--$RH$ space. Any point on this curve satisfies the VPD requirement. The system selects the energy-minimal point by solving:

\begin{align}
\min_{T^*, \text{RH}^*} \; E(T^*, \text{RH}^*) &= \alpha_h \cdot \max(T^* - T_{\text{cur}}, 0) \nonumber \\
&\quad + \alpha_c \cdot \max(T_{\text{cur}} - T^*, 0) \nonumber \\
&\quad + \alpha_d \cdot \max(\text{RH}_{\text{cur}} - \text{RH}^*, 0) \nonumber \\
&\quad + \alpha_m \cdot \max(\text{RH}^* - \text{RH}_{\text{cur}}, 0)
\label{eq:energy_opt}
\end{align}

subject to:

\begin{equation*}
e_s(T^*) \times (1 - \text{RH}^*/100) = \text{VPD}_{\text{target}}
\end{equation*}
\begin{equation*}
T_{\text{min}} \leq T^* \leq T_{\text{max}}, \quad \text{RH}_{\text{min}} \leq \text{RH}^* \leq \text{RH}_{\text{max}}
\end{equation*}

where $\alpha_h$, $\alpha_c$, $\alpha_d$, $\alpha_m$ are energy cost coefficients for heating, cooling, dehumidification, and humidification respectively.

Using the VPD constraint, the problem is parameterized by temperature:

\begin{equation}
\text{RH}^*(T) = 100 \times \left(1 - \frac{\text{VPD}_{\text{target}}}{e_s(T)}\right)
\label{eq:rh_param}
\end{equation}

reducing the optimization to one dimension:

\begin{equation}
\min_{T \in [T_{\text{min}}, T_{\text{max}}]} E(T, \text{RH}^*(T))
\label{eq:1d_opt}
\end{equation}

This 1D optimization is solved at each control cycle by the neural network optimizer, which learns the energy cost surface from operational data. The sensitivity analysis yields $\partial\text{VPD}/\partial T \approx 2$--$3 \times \partial\text{VPD}/\partial\text{RH}$ at typical CEA temperatures (20--30$\,^{\circ}$C), meaning temperature adjustments have a disproportionately larger effect on VPD --- an asymmetry the optimizer exploits for energy-efficient control.

\subsection{Neural Network PID Self-Tuning}

The inner-loop PID controllers use the velocity (incremental) form:

\begin{align}
\Delta u(k) &= K_p(k) \bigl[e(k) - e(k{-}1)\bigr] + K_i(k)\, e(k) \nonumber \\
&\quad + K_d(k) \bigl[e(k) - 2e(k{-}1) + e(k{-}2)\bigr]
\label{eq:pid_velocity}
\end{align}

\begin{equation}
u(k) = u(k{-}1) + \Delta u(k)
\label{eq:pid_update}
\end{equation}

where $e(k) = \text{SP}(k) - \text{PV}(k)$. Anti-windup is implemented via conditional integration when the output saturates~\cite{ref28}.

Initial gain estimation uses the Ziegler--Nichols relay feedback method~\cite{ref28}: a relay of amplitude $\pm h$ is applied to the feedback loop, producing sustained oscillations at the ultimate frequency. The ultimate gain $K_u = 4h/(\pi a)$ and ultimate period $T_u$ are extracted, yielding initial gains $K_p = 0.6\,K_u$, $T_i = T_u/2$, $T_d = T_u/8$. Typical auto-tuning time: 15--30 minutes per control loop.

Online adaptation is performed by a three-layer backpropagation neural network:

\begin{align}
\text{Input: } \mathbf{x}(k) = \bigl[&e(k),\; e(k{-}1),\; e(k{-}2),\; \Delta e(k), \nonumber \\
&\text{SP}(k),\; \text{PV}(k),\; u(k{-}1)\bigr]
\label{eq:nn_input}
\end{align}

The hidden layer consists of 3 neurons with sigmoid activation. The output layer produces 3 neurons:

\begin{align}
\bigl[K_p(k),\; K_i(k),\; K_d(k)\bigr] = \bigl[&\sigma(r_1) \cdot K_{p,\text{max}}, \nonumber \\
&\sigma(r_2) \cdot K_{i,\text{max}}, \nonumber \\
&\sigma(r_3) \cdot K_{d,\text{max}}\bigr]
\label{eq:nn_output}
\end{align}

where $\sigma$ is the sigmoid function and $K_{n,\text{max}}$ defines the feasible gain range. The weight update follows backpropagation with objective $E(k) = \frac{1}{2}[\text{SP}(k) - \text{PV}(k)]^2$, using $\text{sign}(\partial y/\partial u)$ to approximate the unknown plant Jacobian. The static+dynamic decomposition ensures baseline stability: Ziegler--Nichols initial gains provide the stable operating point, while the neural network applies incremental corrections~\cite{ref12, ref13}.

\subsection{Lyapunov Stability Analysis}

For industrial deployment, stability of the adaptive system must be guaranteed. Consider the Lyapunov function candidate:

\begin{equation}
V(k) = \frac{1}{2}\, e(k)^2
\label{eq:lyapunov}
\end{equation}

The change at each step is:

\begin{equation}
\Delta V(k) = V(k{+}1) - V(k) = \frac{1}{2}\bigl[e(k{+}1)^2 - e(k)^2\bigr]
\label{eq:lyapunov_delta}
\end{equation}

The adaptive learning rate $\eta$ is constrained such that $\Delta V(k) < 0$ for all $k$ where $e(k) \neq 0$. This is achieved by bounding the neural network weight updates so that gains $K_p(k)$, $K_i(k)$, $K_d(k)$ remain within the stability region defined by the Ziegler--Nichols initial parameters $\pm \Delta_{\text{max}}$. The L3 autonomy guardrails (Section~IV) provide a physical-domain manifestation of this constraint: the AI may modify setpoints only within proven-safe envelopes.

\subsection{Multi-Objective Optimization}

The overall optimization integrates VPD tracking accuracy, energy cost, and equipment preservation:

\begin{align}
R(s_t, a_t) &= -w_1 \bigl(\text{VPD}_{\text{target}} - \text{VPD}_{\text{measured}}\bigr)^2 \nonumber \\
&\quad - w_2\, E_{\text{cost}}(a_t) - w_3\, W_{\text{equip}}(a_t) \nonumber \\
&\quad - w_4 \sum_i \max\bigl(0,\; g_i(s_t)\bigr)^2
\label{eq:reward}
\end{align}

where $w_1 \ldots w_4$ are weighting coefficients, $E_{\text{cost}}$ represents electricity cost proportional to actuator power, $W_{\text{equip}}$ penalizes rapid actuator switching to reduce mechanical wear (addressing the short-cycling problem with minimum compressor run times of 5~minutes and minimum off times of 3~minutes~\cite{ref1}), and the constraint violation terms $g_i$ enforce temperature, humidity, and CO\textsubscript{2} bounds.

\section{Progressive Autonomy Model}

Adapting the industrial autonomy taxonomy of Gamer~\textit{et~al.}~\cite{ref14} to the specific requirements of CEA operations, we define four graduated autonomy levels (Table~\ref{tab:autonomy}).

\begin{table*}[!t]
\centering
\caption{Progressive Autonomy Levels for CEA Automation}
\label{tab:autonomy}
\begin{tabular}{@{}llll@{}}
\toprule
\textbf{Level} & \textbf{Name} & \textbf{Capability} & \textbf{Action authority} \\
\midrule
L1 & Observation & Anomaly detection & Alerts only \\
L2 & Recommendation & Pattern analysis with confidence scoring & Operator decides \\
L3 & Autonomous (bounded) & Setpoint adjustment within guardrails & Automated with logging \\
L4 & Full optimization & Digital twin simulation + cross-facility learning & Automated, full scope \\
\bottomrule
\end{tabular}
\end{table*}

\subsection{Level L1 --- Anomaly Detection}

The system monitors all sensor streams and detects deviations from learned behavior using an autoencoder-based anomaly detector trained on 14-day rolling baselines. Anomaly categories include: (1)~sensor drift, detected via cross-sensor z-score comparison; (2)~equipment degradation, identified through trending analysis of setpoint achievement time; (3)~environmental anomalies via statistical comparison with rolling baselines; and (4)~irrigation anomalies through dry-back curve analysis. At L1, the system generates alerts but takes no autonomous action.

\subsection{Level L2 --- Recommendations with Confidence Scoring}

The AI module generates specific corrective recommendations, each with a quantified confidence score. For example: ``Reduce night temperature by 1.5$\,^{\circ}$C --- current differential is insufficient for generative growth at week~5 of flowering. Based on 23 analogous cycles. Confidence: 76\%.'' Operators can accept, dismiss, schedule, or request explanations. Dismissed recommendations feed back into the learning system, reducing false positive rates over time.

\subsection{Level L3 --- Autonomous Control within Guardrails}

The system autonomously adjusts parameters within operator-defined bounds (Table~\ref{tab:guardrails}).

\begin{table}[H]
\centering
\caption{L3 Autonomy Guardrails}
\label{tab:guardrails}
\footnotesize
\setlength{\tabcolsep}{3pt}
\begin{tabular}{@{}lll@{}}
\toprule
\textbf{Parameter} & \textbf{Allowed auto-correction} & \textbf{Prohibited actions} \\
\midrule
Temperature & $\pm 2\,^{\circ}$C from recipe & --- \\
Humidity & $\pm 5\%$ from recipe & --- \\
Irrigation volume & $\pm 20\%$ & --- \\
EC & $\pm 0.3$~mS/cm from recipe & --- \\
Photoperiod & --- & All modifications \\
Growth stage transitions & --- & All modifications \\
\bottomrule
\end{tabular}
\end{table}

Photoperiod and stage transitions are prohibited at L3 as they affect plant physiology irreversibly. Every autonomous action is logged with timestamp, parameter, old/new values, reason, and confidence score, providing full auditability and undo capability.

\subsection{Level L4 --- Full Autonomous Optimization}

Activated only after $\geq$3 complete growth cycles with documented outcomes. L4 capabilities include digital twin simulation for recipe testing before application~\cite{ref25}, predictive yield modeling, energy cost optimization through load shifting to off-peak tariff hours, and cross-facility knowledge transfer.

\section{Multi-Facility Deployment Methodology}

\subsection{Standardized Onboarding}

Traditional building management system (BMS) commissioning requires weeks to months of on-site engineering. The IOGRUCloud platform achieves 1--5 day commissioning through standardization of:

\begin{enumerate}
\item \textit{BACnet object tables} --- Pre-built templates for all supported HVAC equipment families (Carrier, Trane, Daikin, LG, Mitsubishi) map standard BACnet objects to the platform's internal data model. When a facility uses supported equipment, integration is configuration rather than custom engineering.

\item \textit{Unified sequences of operations} --- Standard control sequences for common CEA HVAC configurations (split systems, rooftop units, chilled water, VRF) are parameterized and instantiated per facility.

\item \textit{Standardized I/O mapping} --- A consistent mapping framework between physical I/O points and logical control variables eliminates per-facility custom wiring diagrams.

\item \textit{Auto-tuning on first start} --- The Ziegler--Nichols relay feedback procedure (Section~III-C) automatically characterizes each control loop, eliminating manual PID tuning.
\end{enumerate}

\subsection{Multi-Protocol Integration}

The platform's vendor-agnostic integration layer abstracts equipment diversity behind a unified internal API. Table~\ref{tab:protocols} summarizes the protocol coverage.

\begin{table}[H]
\centering
\caption{Protocol Integration Matrix}
\label{tab:protocols}
\footnotesize
\setlength{\tabcolsep}{3pt}
\begin{tabular}{@{}lll@{}}
\toprule
\textbf{Protocol} & \textbf{Equipment category} & \textbf{Typical manufacturers} \\
\midrule
BACnet/IP & HVAC, BAS & Carrier, Trane, Daikin \\
Modbus RTU/TCP & Sensors, VFDs & METER, Gavita, Quest \\
0--10~V analog & LED, VFDs & Fluence, Phantom \\
4--20~mA analog & Transmitters & EC, pH, pressure, flow \\
SDI-12 & Substrate & METER TEROS 12/21 \\
MQTT & IoT devices & Custom gateways \\
OPC~UA & Interoperability & Schneider, Siemens \\
REST/WS & Cloud, apps & Integrations \\
\bottomrule
\end{tabular}
\end{table}

This multi-protocol capability enables the platform to integrate into existing facilities without requiring equipment replacement --- a critical factor for commercial adoption where CEA operators have existing capital investments in diverse equipment.

\subsection{Cross-Facility Transfer Learning}

Operational patterns learned at established facilities are transferred to new deployments through the cloud tier:

\begin{enumerate}
\item \textit{Optimal VPD trajectories} --- Per-crop, per-stage VPD setpoint curves refined from fleet-wide data.
\item \textit{Equipment-specific PID parameters} --- Learned tuning parameters for specific equipment models are transferred to new installations with the same hardware.
\item \textit{Seasonal control templates} --- Climate-zone-specific seasonal strategies (winter heating optimization, summer dehumidification) accelerate adaptation.
\item \textit{Anomaly detection baselines} --- Normal operating envelopes established across the fleet improve anomaly detection at new sites from day one.
\end{enumerate}

\section{Experimental Results}

\subsection{Deployment Scale and Methodology}

The system has been deployed across 30+ commercial CEA facilities in 8 U.S. states spanning diverse climate zones: hot-dry desert (Arizona, Nevada), humid subtropical (Florida), and humid continental (Illinois, Michigan, Ohio, New Jersey, Pennsylvania). Total managed area exceeds 930,000~m\textsuperscript{2} (10,000,000~sq~ft) with 500+ individual control zones. Data collection covers 2017--2024 (7+ years of continuous operation).

The comparative methodology uses a before/after design: each facility's performance under conventional static-setpoint controllers (Priva, Argus, TrolMaster) serves as the baseline, measured for $\geq$3~months prior to IOGRUCloud deployment. Post-deployment metrics are computed over the full operational period.

\subsection{Multi-Facility Aggregate Results}

Table~\ref{tab:aggregate} presents aggregate performance metrics averaged across the deployment fleet.

\begin{table}[H]
\centering
\caption{Average Performance Comparison Across 30+ Facilities (2017--2024)}
\label{tab:aggregate}
\footnotesize
\setlength{\tabcolsep}{3pt}
\begin{tabular}{@{}llll@{}}
\toprule
\textbf{Metric} & \textbf{Before} & \textbf{After} & \textbf{Change} \\
\midrule
HVAC energy & 100\% & 62--70\% & $-$30 to $-$38\% \\
Water use & 100\% & 78--85\% & $-$15 to $-$22\% \\
VPD $\sigma$ (kPa) & 0.15--0.25 & 0.04--0.08 & $-$68 to $-$73\% \\
Recovery time & 12--25~min & 4--8~min & $-$60 to $-$67\% \\
Alarms/month & 8--15 & 1--3 & $-$75 to $-$80\% \\
Downtime & 4.2~h/mo & 1.1~h/mo & $-$74\% \\
\bottomrule
\end{tabular}
\end{table}

The HVAC energy savings of 30--38\% are consistent with published benchmarks: the best real-world result in the literature is 35.44\% from a single-facility RL deployment~\cite{ref29}, while simulation-based studies report up to 57\%~\cite{ref8}. Our results are notable for being achieved across 30+ facilities with heterogeneous equipment, diverse climate zones, and extended temporal scope --- conditions that typically degrade control performance compared to controlled single-site experiments.

The reduction in critical alarms (75--80\%) reflects the system's predictive capability: disturbances are anticipated via weather forecast integration and sensor trend analysis, and preemptively mitigated rather than detected reactively. The 74\% reduction in equipment downtime correlates with the anomaly detection engine (L1) identifying equipment degradation patterns before failure.

\subsection{Case Study 1: Desert Climate (Nevada, 40,000 sq ft)}

Table~\ref{tab:nevada} presents results from a facility operating in Nevada's extreme hot-dry desert climate.

\begin{table}[H]
\centering
\caption{Case Study --- Nevada, 40,000~sq~ft, Desert Climate}
\label{tab:nevada}
\footnotesize
\setlength{\tabcolsep}{3pt}
\begin{tabular}{@{}llll@{}}
\toprule
\textbf{Parameter} & \textbf{Before} & \textbf{After} & \textbf{Change} \\
\midrule
Electricity & \$18,400/mo & \$11,700/mo & $-$36.4\% \\
Water & \$3,200/mo & \$2,450/mo & $-$23.4\% \\
VPD $\sigma$ & 0.22~kPa & 0.06~kPa & $-$72.7\% \\
Yield (g/sq~ft) & 52 & 61 & +17.3\% \\
Energy/yield & 1.42~kWh/g & 0.97~kWh/g & $-$31.7\% \\
\bottomrule
\end{tabular}
\end{table}

In this environment with extreme diurnal temperature swings ($>$25$\,^{\circ}$C day-night differential) and very low ambient humidity ($<$15\% RH), the cascading VPD control provides particular advantage. Rather than maintaining rigid temperature and humidity setpoints --- requiring maximum HVAC capacity during afternoon peaks --- the AI optimizer exploits the VPD constraint surface. For example, allowing temperature to rise to 25.5$\,^{\circ}$C while adjusting humidity to 52\% achieves the same target VPD as the conventional 24$\,^{\circ}$C/55\% setpoint, but with significantly reduced cooling load.

Predictive control based on weather forecast integration~\cite{ref30} further reduces consumption: before anticipated heat waves, the system pre-cools during nighttime off-peak hours, exploiting the desert's cool nights and lower electricity tariffs.

\subsection{Case Study 2: Continental Climate (Illinois, 120,000 sq ft)}

Table~\ref{tab:illinois} presents results from a larger facility in Illinois with extreme seasonal variation.

\begin{table}[H]
\centering
\caption{Case Study --- Illinois, 120,000~sq~ft, Continental Climate}
\label{tab:illinois}
\footnotesize
\setlength{\tabcolsep}{3pt}
\begin{tabular}{@{}llll@{}}
\toprule
\textbf{Parameter} & \textbf{Before} & \textbf{After} & \textbf{Change} \\
\midrule
HVAC (winter) & \$42,000/mo & \$27,300/mo & $-$35.0\% \\
HVAC (summer) & \$31,000/mo & \$21,400/mo & $-$31.0\% \\
Operator hrs/mo & 480~h & 80~h & $-$83.3\% \\
Crop loss/year & 3 & 0 & $-$100\% \\
System ROI & --- & 7.2~months & --- \\
\bottomrule
\end{tabular}
\end{table}

The continental climate presents distinct challenges: extreme winter cold requires managing heating costs while preventing condensation; hot humid summers stress dehumidification capacity. The system's adaptive behavior is particularly evident during seasonal transitions --- automatically adjusting control strategies as outdoor conditions shift, whereas the conventional Argus system required manual seasonal reconfiguration.

The 83.3\% reduction in monitoring labor reflects the progression from L1 to L3 autonomy. Operators shifted from continuous monitoring to exception-based supervision.

\subsection{Comparison with Published Literature}

Table~\ref{tab:comparison} contextualizes the deployment scale against the existing body of work.

\begin{table*}[!t]
\centering
\caption{Real-World Deployment Scale Comparison}
\label{tab:comparison}
\begin{tabular}{@{}lllll@{}}
\toprule
\textbf{Study} & \textbf{Facilities} & \textbf{Duration} & \textbf{Domain} & \textbf{Control type} \\
\midrule
All field experiments combined~\cite{ref19} & $\sim$20 buildings & 43 days total & HVAC & Various AI \\
Moshari \textit{et~al.}~\cite{ref21} & 13 buildings & 138 days & District heating & RL \\
FarmBeats~\cite{ref22} & 2 farms & 6 months & Open-field ag & Monitoring only \\
Cao \textit{et~al.} (iGrow)~\cite{ref11} & 1 greenhouse & $\sim$1 season & CEA & RL \\
Hemming \textit{et~al.}~\cite{ref31} & 1 greenhouse & 1 season & CEA & AI-assisted \\
\textbf{IOGRUCloud (this work)} & \textbf{30+ facilities} & \textbf{7+ years} & \textbf{CEA} & \textbf{NN-PID + VPD cascade} \\
\bottomrule
\end{tabular}
\end{table*}

The difference is not incremental --- it represents orders of magnitude in both facility count and operational duration. This scale enables observations impossible in short-term deployments: long-term equipment degradation detection, multi-season adaptation patterns, and statistically significant cross-facility comparisons.

\section{Deployment Lessons}

Seven years of production operation across 30+ facilities with 50+ equipment manufacturers have yielded practical lessons that are absent from the simulation-focused literature.

\subsection{What Worked}

\begin{enumerate}
\item \textit{VPD as the master setpoint eliminated cross-coupling conflicts.} The most immediate and consistent improvement across all facilities was the elimination of simultaneous heating and dehumidification --- a common waste pattern in conventional controllers where independent $T$ and $RH$ loops fight each other.

\item \textit{Edge-first architecture was essential for reliability.} Network outages, ISP failures, and cloud service disruptions occur frequently in commercial facilities. Facilities that lost internet connectivity for hours or days continued operating normally because all control logic executed locally. The cloud tier proved valuable for optimization but not for real-time control.

\item \textit{Standardized BACnet object tables enabled rapid commissioning.} The pre-built equipment templates reduced commissioning from weeks to 1--5 days. The most significant time savings came from eliminating per-facility custom BACnet point mapping.

\item \textit{Progressive autonomy built operator trust.} Starting at L1 (observation only) and gradually increasing autonomy over weeks allowed operators to build confidence in the system. Facilities that attempted to jump directly to L3 autonomous control experienced higher resistance and more frequent manual overrides.

\item \textit{Cross-facility transfer learning accelerated new deployments.} New facilities with equipment models previously seen in the fleet achieved optimal performance 60--70\% faster than early deployments without transfer learning baselines.
\end{enumerate}

\subsection{What Failed or Required Iteration}

\begin{enumerate}
\item \textit{Sensor drift at scale was the dominant operational challenge.} Across 30+ facilities with thousands of sensors, sensor drift --- not control algorithm performance --- was the primary source of control degradation. The median filtering redundancy (Section~II-A) was added after the first year of operation when drift-related false alarms became the leading maintenance issue.

\item \textit{Generic PID auto-tuning was insufficient for some HVAC configurations.} The Ziegler--Nichols relay feedback method produced acceptable initial gains for approximately 80\% of control loops. The remaining 20\% --- primarily large chilled water systems with significant transport delay --- required Cohen--Coon identification or manual refinement. The neural network adaptation eventually compensated, but initial performance suffered during the learning period.

\item \textit{Facility operators initially resisted automated CO\textsubscript{2} management.} CO\textsubscript{2} injection coordination with ventilation cycles required careful cross-subsystem logic. Early deployments paused CO\textsubscript{2} injection during ventilation events but did not account for room volume and air exchange rate, leading to suboptimal CO\textsubscript{2} recovery times. The system was refined to model room volume dynamics.

\item \textit{Legacy equipment integration remained the most time-consuming commissioning task.} While standardized BACnet templates worked well for modern HVAC equipment, older installations with proprietary serial protocols or analog-only interfaces required custom gateway configuration that extended commissioning beyond the 5-day target.

\item \textit{Energy cost coefficients required seasonal recalibration.} The energy cost parameters ($\alpha_h$, $\alpha_c$, $\alpha_d$, $\alpha_m$) in the VPD optimization~(\ref{eq:energy_opt}) required quarterly updates to reflect changing utility tariff structures and seasonal equipment efficiency curves. Automating this calibration from utility bill data was added in the third year of operation.
\end{enumerate}

\subsection{Scalability Observations}

\begin{enumerate}
\item \textit{Linear scaling of sensor data, sub-linear scaling of control complexity.} Adding zones to a facility scales sensor data linearly, but control complexity grows sub-linearly because adjacent zones share environmental coupling --- the system exploits this coupling for coordinated control.

\item \textit{Fleet diversity improved model robustness.} Counter-intuitively, deploying across diverse climate zones and equipment configurations produced more robust models than deployments across identical facilities. The diversity forced the neural network to learn generalizable patterns rather than overfitting to specific equipment characteristics.

\item \textit{99.9\% uptime achieved through redundancy, not reliability.} Individual components failed regularly. The 99.9\% system uptime target was achieved through edge-level failsafe modes, sensor redundancy, and graceful degradation --- not through preventing failures.
\end{enumerate}

\section{Conclusion}

This paper presented IOGRUCloud, a three-tier IoT platform for AI-driven climate control in Controlled Environment Agriculture, deployed across 30+ commercial facilities in 8 U.S. climate zones over 7+ years. The principal contributions are:

\begin{enumerate}
\item \textit{Cascading VPD control.} Elevating VPD from a monitored metric to the primary cascading setpoint, with neural network-optimized energy-minimal $T$--$RH$ decomposition, achieves 30--38\% HVAC energy savings and 68--73\% VPD stability improvement.

\item \textit{Neural network PID self-tuning.} A 7-3-3 backpropagation network with Lyapunov stability guarantees adapts PID gains online --- the first reported production deployment in CEA.

\item \textit{Progressive autonomy (L1--L4).} A domain-specific adaptation of industrial autonomy concepts enables graduated trust-building with operator guardrails.

\item \textit{Scale of validation.} 30+ facilities, 930,000+ m\textsuperscript{2}, 500+ control zones, 7+ years --- exceeding the combined duration of all published AI-based HVAC field experiments by approximately 60$\times$.

\item \textit{Practical deployment lessons.} VPD-based cascade control, edge-first architecture, and standardized commissioning are identified as the most impactful design decisions; sensor drift management and legacy equipment integration are identified as the dominant operational challenges.
\end{enumerate}

Future work includes federated learning for cross-facility knowledge transfer without exchanging raw data, integration of dynamic electricity pricing for real-time load optimization, extension to adjacent domains (data center cooling, pharmaceutical cleanrooms), and open-sourcing the standardized BACnet equipment templates.

\section*{Acknowledgment}

The author would like to thank the cultivation teams and facility operators across all deployed sites for their collaboration and operational data. The author also acknowledges the IOGRU engineering team for their contributions to platform development and deployment.

During the preparation of this manuscript, generative AI tools (Claude, Anthropic) were used for literature search assistance and language editing. All technical content, system architecture descriptions, mathematical formulations, experimental data, and conclusions are the sole work of the author. The author reviewed and edited all AI-assisted content and takes full responsibility for the final manuscript.


\begin{IEEEbiography}[{\includegraphics[width=1in,height=1.25in,clip,keepaspectratio]{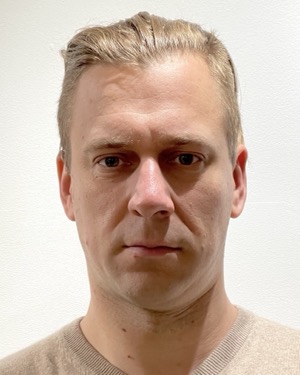}}]{ANDRII VAKHNOVSKYI}
received the M.Sc. degree in Computer Engineering and System Programming from the National Technical University ``Kharkiv Polytechnic Institute'' (NTU ``KhPI''), Kharkiv, Ukraine, in 2011. He is the Founder and CTO of IOGRU LLC, New York, NY, USA, where he designed and deployed the IOGRUCloud platform across 30+ commercial CEA facilities in 8 U.S. states. His research interests include industrial IoT, adaptive control systems, neural network-augmented PID control, and energy-efficient automation for controlled environment agriculture. Previously, he worked on building management systems with Schneider Electric and Siemens authorized integrators across 20+ industrial projects including data centers, power generation facilities, and hospitals.
\end{IEEEbiography}

\end{document}